\documentclass[conference]{IEEEtran}
\IEEEoverridecommandlockouts
\def\BibTeX{{\rm B\kern-.05em{\sc i\kern-.025em b}\kern-.08em T\kern-.1667em\lower.7ex\hbox{E}\kern-.125emX}}
\usepackage{graphicx} 
\usepackage{amsmath,amssymb}
\usepackage{amsfonts}
\usepackage{bm}
\usepackage{enumitem}
\usepackage{acronym}
\usepackage{cite}

\usepackage{multirow}
\usepackage{microtype}
\usepackage{balance}
\usepackage{glossaries}
\usepackage[utf8]{inputenc}
\usepackage{multirow}
\usepackage{hhline}
\usepackage{colortbl}
\usepackage{multicol}
\usepackage[table,xcdraw]{xcolor}  
\usepackage{diagbox}  
\usepackage{booktabs} 
\usepackage{makecell} 
\usepackage{graphicx}    
\usepackage[font=footnotesize,labelformat=simple]{subcaption} 
\usepackage{comment} 

\usepackage[a-1b]{pdfx}

\usepackage[font = footnotesize, textfont=up]{caption}
\newacronym{cpp}{CPP}{carrier phase positioning}
\newacronym{cnn}{CNN}{convolutional neural network}

\newacronym{bie}{BIE}{best integer equivariant}
\newacronym{tof}{ToF}{time-of-flight}
\newacronym{toa}{ToA}{time-of-arrival}
\newacronym{tdoa}{TDoA}{time-difference-of-arrival}
\newacronym{psd}{PSD}{power spectral density}
\newacronym{csi}{CSI}{channel state information}
\newacronym{snr}{SNR}{signal-to-noise ratio}
\newacronym{nn}{NN}{neural network}
\newacronym{los}{LoS}{line-of-sight}
\newacronym{cf}{CF}{cell-free}
\newacronym{ml}{ML}{machine learning}
\newacronym{ue}{UE}{user equipment}
\newacronym{mimo}{MIMO}{multiple-input multiple-output}
\newacronym{mle}{MLE}{maximum likelihood estimation}
\newacronym{dl}{DL}{deep learning}
\newacronym{ap}{AP}{antenna point}
\newacronym{flop}{FLOP}{floating-point operation}
\newacronym{dnn}{DNN}{deep neural network}
\newacronym{mlp}{MLP}{multi-layer perceptron }
\newacronym{rmse}{RMSE}{Root mean-squared error}
\newacronym{relu}{ReLU}{Rectified Linear Unit}
\newacronym{conv2d}{Conv2D}{2D-Convolutional}
\newacronym{mse}{MSE}{mean-squared error}
\newacronym{ecdf}{ECDF}{empirical cumulative distribution function}
\newacronym{srs}{SRS}{sounding reference signal}
\newacronym{gd}{GD}{gradient descent}
\newacronym{3gpp}{3GPP}{3rd Generation Partnership Project}
\newacronym{5gnr}{5G NR}{5G New Radio}
\newacronym{prs}{PRS}{positioning reference signal}
\newacronym{gnss}{GNSS}{global navigation satellite system}
\newacronym{dmrs}{DMRS}{demodulation reference signal}
\newacronym{nlos}{NLoS}{non-line-of-sight}

\def\BibTeX{{\rm B\kern-.05em{\sc i\kern-.025em b}\kern-.08em
    T\kern-.1667em\lower.7ex\hbox{E}\kern-.125emX}}
\begin{document}
\bstctlcite{IEEEexample:BSTcontrol}

\title{Failure Tolerant Phase-Only Indoor Positioning via Deep Learning
}

\author{\IEEEauthorblockN{Fatih Ayten\IEEEauthorrefmark{1}, Mehmet C. Ilter\IEEEauthorrefmark{1}, Akshay Jain\IEEEauthorrefmark{2}, Ossi Kaltiokallio\IEEEauthorrefmark{1}, Jukka Talvitie\IEEEauthorrefmark{1}, \\ 
Elena Simona Lohan\IEEEauthorrefmark{1}, Henk Wymeersch\IEEEauthorrefmark{3}, and Mikko Valkama\IEEEauthorrefmark{1} }  
\vspace{2mm}
\IEEEauthorblockA{
\IEEEauthorrefmark{1}Electrical Engineering Unit, Tampere Wireless Research Center, Tampere University, Finland\\
\IEEEauthorrefmark{2}Radio Systems Research, Nokia Bell Labs, Espoo, Finland\\
\IEEEauthorrefmark{3}Department of Electrical Engineering, Chalmers University of Technology, Sweden\\
Email: fatih.ayten@tuni.fi, mehmet.ilter@tuni.fi, akshay.2.jain@nokia-bell-labs.com, ossi.kaltiokallio@tuni.fi, \\ jukka.talvitie@tuni.fi, elena-simona.lohan@tuni.fi, henkw@chalmers.se, mikko.valkama@tuni.fi
}}

\maketitle

\begin{abstract}
High-precision localization turns into a crucial added value and asset for next-generation wireless systems. \Gls{cpp} enables sub-meter to centimeter-level accuracy and is gaining interest in 5G-Advanced standardization. While \gls{cpp} typically complements \gls{toa} measurements, recent literature has introduced a \emph{phase-only} positioning approach in a distributed antenna/MIMO system context with minimal bandwidth requirements, using \gls{dl} when operating under ideal hardware assumptions. In more practical scenarios, however, antenna failures can largely degrade the performance. In this paper, we address the challenging phase-only positioning task, and propose a new \gls{dl}-based localization approach harnessing the so-called hyperbola intersection principle, clearly outperforming the previous methods. Additionally, we consider and propose a processing and learning mechanism that is robust to antenna element failures. Our results show that the proposed \gls{dl} model achieves robust and accurate positioning despite antenna impairments, demonstrating the viability of data-driven, impairment-tolerant phase-only positioning mechanisms. Comprehensive set of numerical results demonstrates large improvements in localization accuracy against the prior art methods.
\end{abstract}


\glsresetall    

\section{Introduction}
The \gls{3gpp} Release 17 and 18 standards have introduced stringent positioning requirements, targeting centimeter-level accuracy \cite{3gpp_tr_38_859_2024,Cha2025}. As a result, high-precision positioning has become a key feature in next-generation terrestrial mobile communication systems. \emph{\Gls{cpp}} is well known for enabling precise ranging, achieving sub-meter to centimeter-level accuracy \cite{10475845}. While \gls{cpp} techniques have been widely employed in the \gls{gnss} context \cite{ding2021carrier}, their adoption in \gls{3gpp} standardization has gained traction, particularly within 5G-Advanced, where carrier phase measurements are being explored as an emerging technology to enhance positioning capabilities \cite{10536135}. To this end, the \gls{cpp} mechanisms have been investigated for 5G networks using different reference signals, such as \gls{prs} \cite{5g_carrier_phase_prs_1,5g_carrier_phase_prs_2} and \gls{dmrs} \cite{5g_carrier_phase_dmrs}, further demonstrating their potential in cellular-based positioning. Additionally, extracting highly accurate situational awareness is also expected to be a key asset in future 6G networks \cite{10536135}.

\begin{figure}[t!] 
    \centering
    \vspace{-0.5mm}
    \includegraphics[width=0.38\textwidth]{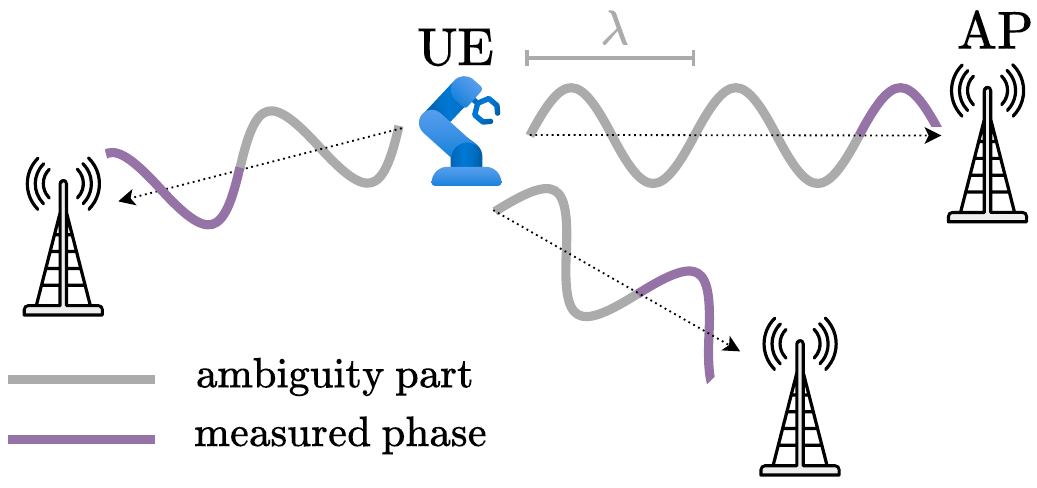}
    \caption{Illustration of uplink UE positioning with distributed phase-coherent antenna points (APs) where only the carrier phase measurements at different APs are used to estimate the UE position.}\vspace{2mm}
    \label{fig:system_model}
\end{figure}

In general, the available \gls{cpp} techniques mostly utilize phase measurements as an additional asset or measurement, complementing the more ordinary \gls{toa} or \gls{tdoa} positioning measurements \cite{10437902, Cha2025, 10475845, 5g_carrier_phase_prs_1}. 
Alternatively, \cite{ayten2025phase} has recently proposed a \emph{phase-only} \gls{ue} positioning approach in distributed phase-coherent \gls{mimo} systems by introducing two \gls{dl} solutions that use only uplink carrier phase data while operating under a perfect hardware assumption. This scenario is conceptually illustrated in Fig.~\ref{fig:system_model}, and also forms the main overall scope of this work.

In practical implementations, antenna elements are vulnerable to failures caused by hardware aging, environmental conditions, or catastrophic events \cite{10437177}. Beyond hardware impairments, as observed particularly in multi-antenna \gls{gnss} receivers, spoofing attacks can introduce additional performance-degrading impairments \cite{10719606,harvanek2024survey}. In such scenarios, data-driven learning-based methods offer a promising alternative by enabling systems to learn robust representations directly from observed data, without relying on explicit system modeling. These methods have shown increasing potential in adapting to complex and imperfect conditions \cite{8054694,9690703,e2e_radar_TAES_2022,spatial_mmWave_Pos_E2E_2023}. While early studies focused on communication-oriented tasks \cite{8054694}, recent work demonstrates the applicability of data-driven techniques to sensing and localization, where resilience to different types of hardware and other impairments is essential \cite{e2e_radar_TAES_2022}.

In this paper, we address the challenging phase-only positioning task, via deep learning, with the primary purpose of improving the localization accuracy beyond that offered by the existing state-of-the-art approach in \cite{ayten2025phase}.
Furthermore, motivated by the above practical impairment considerations, we investigate and extend the phase-only positioning approach 
to scenarios involving antenna element failures. Specifically, we propose and present a \gls{nn} model trained on data from both fault-free and antenna failure conditions to improve robustness against hardware impairments. Additionally, we employ a novel, so-called hyperbola intersection method, leveraging phase differences as the primary input modality for position estimation. Our numerical results show that the proposed method achieves sub-centimeter accuracy, with 95th percentile errors consistently below 2 cm across various transmit powers and failure probabilities in the considered evaluation environment. Additionally, we demonstrate that including failure scenarios during training significantly improves the considered \gls{nn} model’s ability to adapt to antenna faults, enabling reliable and precise positioning even under practical failure conditions. These results validate the effectiveness of the proposed learning-based phase-only positioning framework in real-world, imperfect distributed antenna environments.\vspace{-3mm}

\section{System Model}
\vspace{-2mm}
Similar to \cite{ayten2025phase}, we consider an uplink scenario involving a \gls{ue} and $I$ distributed, mutually phase-synchronized \glspl{ap}. Fig.~\ref{fig:system_model} depicts an example scenario involving three APs such that $I=3$. The \gls{ue} has an unknown position denoted by $\boldsymbol{x}_{\text{ue}} \in \mathbb{R}^2$, while the positions of the \glspl{ap} are known and represented as $\boldsymbol{x}_{\text{ap,}i} \in \mathbb{R}^2$ for $i \in \{0, \dots, I-1\}$. For simplicity and brevity, the analysis is restricted to \gls{los} propagation conditions, while the extensions to multipath scenarios and three-dimensional space, i.e., $\mathbb{R}^3$ form an important topic for our future work.

The \gls{ue} transmits a unit-power narrowband pilot symbol, denoted as $s$, with a transmit power $P_T$ where its bandwidth is equal to $W$. The received signal at the $i$-th \gls{ap} over the \gls{los} path is given by
\begin{equation}\label{eq:rx_signal}
y_i = f_i \sqrt{P_T/W} \rho_i \exp\left(-j\left(\frac{2\pi}{\lambda}d_i - \phi_{\text{ue}}\right)\right)s + v_i, 
\end{equation}
where $\rho_i$ denotes the path loss between the \gls{ue} and the $i$-th \gls{ap}, $\lambda$ is the wavelength, $d_i =  \Vert\boldsymbol{x}_{\text{ue}}-\boldsymbol{x}_{\text{ap,}i}\Vert$ is the Euclidean distance between the \gls{ue} and the $i$-th \gls{ap}, $\phi_{\text{ue}}$ represents the common phase offset between the \gls{ue} and the \gls{ap} network, and $v_i \sim \mathcal{CN}(0, N_0)$ denotes additive complex Gaussian noise. The binary indicator $f_i \in \{0,1\}$ models hardware-related failures at the $i$-th \gls{ap}, where $f_i = 1$ indicates operational status and $f_i = 0$ represents failure. We assume that each \gls{ap} fails independently with probability $p_f$. By processing $y_i$, the resulting phase observation can be expressed as \\
\begin{equation}\label{eq:phase_observations}
    \theta_i = f_i \times (-\frac{2\pi}{\lambda}d_i+\phi_{\text{ue}}+ 2 \pi z_i)  + n_i,
\end{equation}
where $z_i \in \mathbb{Z}$ is the integer ambiguity, and $n_i \sim \mathcal{N}(0, \sigma_i^2)$ models phase measurement noise. Using \eqref{eq:phase_observations}, one specific AP is assigned as a reference unit by the network, e.g., $i = 0$, and we define the differential measurements for $m \in \{1, \dots, I-1\}$ as
\begin{align}
\delta_m &=  -\frac{\lambda}{2\pi}(\theta_m - \theta_0) \label{eq:phase_diff} \\
&= \Delta d_m - \lambda\Delta z_m  - \frac{\lambda \phi_{\text{ue}}}{2\pi}\Delta f_m + \Delta n_m,
\label{eq:diff_meas}
\end{align}
where $\Delta d_m = f_m d_m - f_0 d_0$, $\Delta z_m = f_mz_m - f_0z_0$, $\Delta f_m = f_m - f_0$ and $\Delta n_m = (\lambda / 2\pi)(n_0 - n_m)$. Also, we define the failure-independent counterpart of $\Delta d_m$ and $\Delta z_m$ as $\Delta\overline{d}_m = d_m - d_0$ and $\Delta\overline{z}_m = z_m - z_0$, respectively. Stacking the differential measurements in \eqref{eq:diff_meas}, we obtain the vector  $\boldsymbol{\delta} =\left[\delta_1, \dots , \delta_{I-1}\right]^\top$.

When $f_m=f_0=1$, the term $\delta_m$ geometrically defines a collection of hyperbolas parameterized by $\Delta z_m$, where the foci are located at $(\boldsymbol{x}_{\text{ap,}m},\boldsymbol{x}_{\text{ap,}0})$. Failure of the reference \gls{ap} compromises all elements of the differential measurements, thereby increasing the difficulty of the positioning task.

\section{Proposed Hyperbola Intersection Method}\label{sec:proposed_approach}
In this section, we present the proposed hyperbola intersection method, which integrates \gls{dl} and \gls{gd} techniques to estimate the position of the \gls{ue} using the above phase-based measurements while also detecting \gls{ap} failures. The overall framework is illustrated in Fig.~\ref{fig:proposed_approach} and consists of two main components: a differential ambiguity estimator based on a \gls{mlp}, and a \gls{gd}-based solver. The GD-based solver outputs an estimated \gls{ue} position denoted by $\hat{\boldsymbol{x}}_{\text{ue}}$, along with the final cost value $E_T.$ The computed cost value $E_T$ serves as a confidence level for the position estimate $\hat{\boldsymbol{x}}_{\text{ue}}$. If the cost satisfies $E_T \leq\tau,$ where $\tau$ is a predefined threshold, the framework flags \emph{No \gls{ap} Failure}, indicating the position estimation is considered reliable. Otherwise, it flags a \emph{Potential \gls{ap} Failure}, detecting a potential failure in one or more \glspl{ap}.

\subsection{NN-based Differential Ambiguity Estimator}\label{subsec:nn_model}
A key challenge to achieving reliable \gls{cpp} lies in the resolution of \emph{integer ambiguities} \cite{10448267}. To address this issue, an \gls{mlp}-based neural network model is introduced to generate a vector of failure-independent differential ambiguity estimates 
$\Delta\hat{\boldsymbol{z}} = \left[\Delta\hat{z}_{j_1}, \dots, \Delta\hat{z}_{j_{\lvert\mathcal{J}\rvert}} \right]^\top$ from the differential measurements $\boldsymbol{\delta}$, where $\mathcal{J} = \{ j_k \}_{k=1}^{\lvert\mathcal{J}\rvert} \subseteq \{1, 2, \dots, I-1\}$ denotes the set of AP indices whose failure-independent differential ambiguities are estimated, with the cardinality of $\lvert\mathcal{J}\rvert$. Here, $\Delta\hat{z}_{j_k}$ denotes the estimate of $\Delta \overline{z}_{j_k}$.

Each failure-independent differential ambiguity $\Delta \overline{z}_{j_k}$ is constrained within the geometric bounds $\left[-q_{j_k}-1,q_{j_k} \right] = \left[ \lfloor - \Vert \boldsymbol{x}_{\text{ap,}{j_k}}-\boldsymbol{x}_{\text{ap,}0}\Vert/\lambda \rfloor , \lfloor \Vert\boldsymbol{x}_{\text{ap,}{j_k}}-\boldsymbol{x}_{\text{ap,}0}\Vert/\lambda \rfloor \right]$, where $\lfloor \cdot \rfloor$ represents the floor function, which yields the greatest integer less than or equal to the input. Consequently, each differential ambiguity can take $Q_{j_k} = 2 q_{j_k} +2$ possible labels, and the combined number of possible labels for the differential ambiguities in the set $\mathcal{J}$ is given by $Q=\sum_{k=1}^{\lvert \mathcal{J} \rvert} Q_{j_k}$. 

\begin{figure}[t!] 
    \centering
    \includegraphics[width=0.47\textwidth]{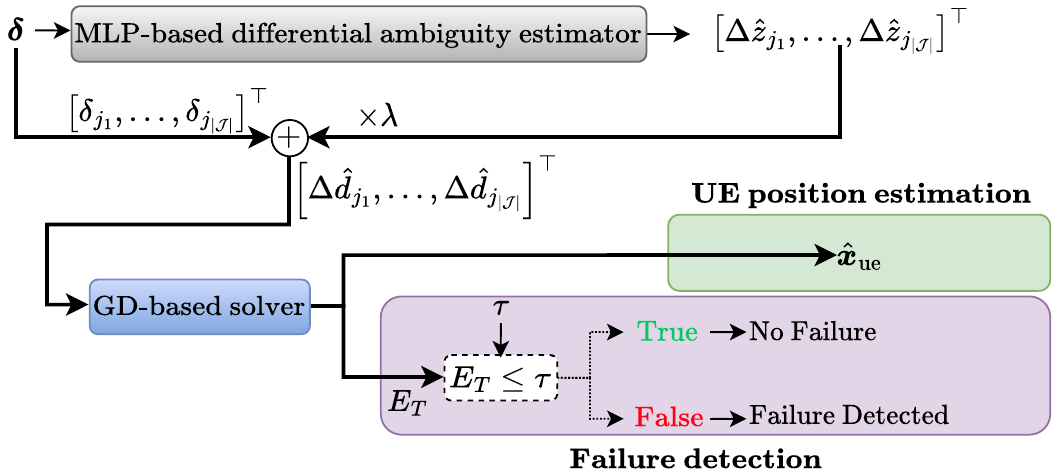}
    \caption{Framework of the proposed approach.}\vspace{-0em}
    \label{fig:proposed_approach}
\end{figure}

\begin{figure}
    \centering
    \vspace{-0em}
    \includegraphics[width=0.3\textwidth]{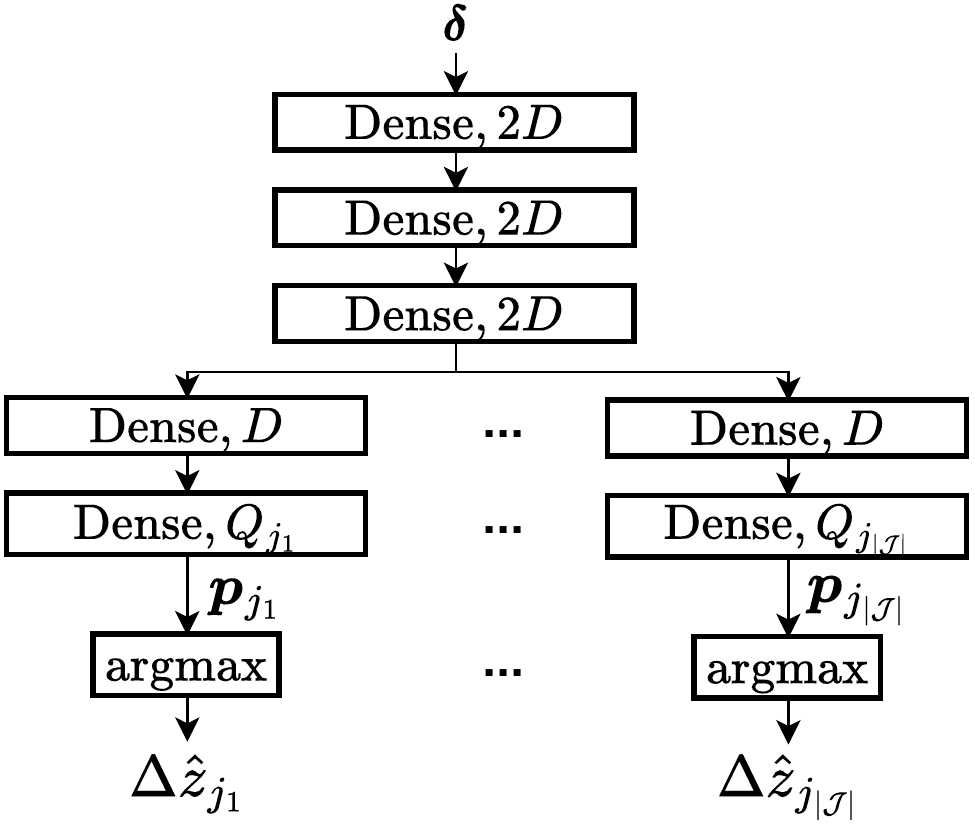}
    \caption{Architecture of the proposed \gls{mlp}-based ambiguity estimation approach: Each rectangle denotes a dense (fully connected) layer, with the number inside indicating the neuron count and each branch outputs a failure-independent differential ambiguity estimation.}\vspace{-0em}
    \label{fig:ambiguity_estimator_figure}
\end{figure}

The architecture of the proposed \gls{mlp} model is illustrated in Fig.~\ref{fig:ambiguity_estimator_figure}. The input is first processed by shared layers, followed by $\lvert \mathcal{J} \rvert$ parallel branches. Both the shared layers and the initial layer of each parallel branch employ the \gls{relu} activation function. The second layer of each branch uses the softmax activation function to produce a probability distribution $\boldsymbol{p}_{j_k} = \left[ p_{j_k,-q_{j_k}-1} , \dots, p_{j_k,q_{j_k}} \right]^\top$,
where each element $p_{j_k,l} \in [0,1]$ represents the probability of $\Delta \overline{z}_{j_k}$ taking the integer value $l$, satisfying $\sum_l p_{j_k,l}=1$. To obtain the final integer ambiguity prediction, each probability distribution is passed through an argmax function which selects the most probable class for each differential ambiguity. This results in the estimate $\Delta\hat{z}_{j_k} = \arg\max_{l \in [-q_{j_k}-1, q_{j_k}]} p_{{j_k},l}$. The sparse categorical cross-entropy \cite{scce_ref} is employed as the loss function to train the \gls{mlp} model. By noting that $p_{j_k,\Delta \overline{z}_{j_k}}$ is the generated probability of the correct label for $\overline{z}_{j_k}$, the loss function can be expressed as
\begin{equation}
\mathcal{L}_{\text{SCCE}} = -\frac{1}{\lvert \mathcal{J} \rvert} \sum_{k=1}^{\lvert \mathcal{J} \rvert} \ln(p_{j_k,\Delta \overline{z}_{j_k}}),
\end{equation}
where $\ln( \cdot)$ is the natural logarithm. Finally, as illustrated in Fig.~\ref{fig:proposed_approach}, the estimated ambiguities, scaled with $\lambda$, are added to the corresponding differential phase measurements to obtain the failure-independent differential distance estimations
\begin{equation}
\begin{split}
\left[\Delta\hat{d}_{j_1}, \dots, \Delta\hat{d}_{j_{\lvert \mathcal{J} \rvert}}\right]^\top &= \left[\delta_{j_1}, \dots , \delta_{j_{\lvert \mathcal{J} \rvert}}\right]^\top \\ 
&\quad + \lambda \left[\Delta\hat{z}_{j_1}, \dots, \Delta\hat{z}_{j_{\lvert \mathcal{J} \rvert}}\right]^\top.
\end{split}
\label{eq:diff_dist_est}
\end{equation}
 
\subsection{Gradient Descent-Based Solver}\label{subsec:grad_model}
We propose an iterative \gls{gd}-based optimization algorithm for joint UE positioning and AP failure detection. The algorithm is initialized with a random initial guess $\boldsymbol{x}_{\text{ue},0}^{\text{(GD)}}$ for the UE position. At each iteration $t \in \{1, 2, \dots, T\}$, the Euclidean distance from the current UE position estimate $\boldsymbol{x}_{\text{ue},t}^{\text{(GD)}}$ to the $j_k$-th \gls{ap} is computed as $d_{j_k,t}^{\text{(GD)}}=\Vert\boldsymbol{x}^{\text{(GD)}}_{\text{ue},t}-\boldsymbol{x}_{\text{ap,}j_k}\Vert$. Also, we compute the Euclidean distance from the current UE position estimate to the reference \gls{ap} as $d_{0,t}^{\text{(GD)}}=\Vert\boldsymbol{x}^{\text{(GD)}}_{\text{ue},t}-\boldsymbol{x}_{\text{ap,}0}\Vert$. Differential distances with respect to the reference \gls{ap} are then formed as $\Delta d_{j_k,t}^{\text{(GD)}}=d_{j_k,t}^{\text{(GD)}}-d_{0,t}^{\text{(GD)}}$. These are compared with the differential distances $\Delta \hat{d}_{j_k}$ predicted by the \gls{mlp}-based ambiguity estimator, which remains fixed during iterations. The residual errors are computed as $e_{j_k,t} = \Delta d_{j_k,t}^{\text{(GD)}}-\Delta \hat{d}_{j_k}$. These residuals are used to define the quadratic cost function $E_t=\sum_{k=1}^{\lvert \mathcal{J} \rvert} (e_{j_k,t})^2$. The position estimate is updated according to $\boldsymbol{x}_{\text{ue},t+1}^{\text{(GD)}} = \boldsymbol{x}_{\text{ue},t}^{\text{(GD)}}-\alpha \boldsymbol{\nabla} E_t$, where $\alpha$ is the learning rate and $\boldsymbol{\nabla} E_t$ denotes the gradient of the cost function with respect to $\boldsymbol{x}_{\text{ue},t}^{\text{(GD)}}$, given by
\begin{equation}\label{eq:gradient}
\boldsymbol{\nabla} E_t = 2\sum_{k=1}^{\lvert \mathcal{J} \rvert} e_{j_k,t} \left(\frac{\boldsymbol{x}_{\text{ue},t}^{\text{(GD)}} - \boldsymbol{x}_{\text{ap},j_k}}{d_{j_k,t}^{\text{(GD)}}} - \frac{\boldsymbol{x}_{\text{ue},t}^{\text{(GD)}} - \boldsymbol{x}_{\text{ap},0}}{d_{0,t}^{\text{(GD)}}}\right).
\end{equation}

After completing $T$ iterations, the final \gls{ue} position estimation $\hat{\boldsymbol{x}}_{\text{ue}} = \boldsymbol{x}_{\text{ue},T}^{\text{(GD)}}$ and the final cost $E_T$ are generated. The framework generates a \emph{No \gls{ap} Failure} flag when $E_T \leq \tau$, which indicates that the estimated position $\hat{\boldsymbol{x}}_{\text{ue}}$ can be considered reliable. Conversely, if $E_T > \tau$, the framework outputs a \emph{Potential \gls{ap} Failure} flag, indicating either an incorrect estimation of differential ambiguities by the \gls{mlp}-based ambiguity estimator or a potential \gls{ap} failure. In this work, we interpret the threshold test primarily as an \gls{ap} failure detection mechanism; thus, $E_T > \tau$ is taken as an indication of a potential \gls{ap} failure. It is important to note that the proposed solver always outputs a \gls{ue} position estimate $\hat{\boldsymbol{x}}_{\text{ue}}$, regardless of the final cost~$E_T$; that is, no position estimate is discarded, even when $E_T > \tau$. The choice of $\tau$ critically affects the failure detection performance: a smaller $\tau$ improves sensitivity to potential \gls{ap} failures but may erroneously reject reliable estimates, while a larger $\tau$ increases tolerance to estimation errors at the risk of missing actual failures.

\section{Inference Complexity}\label{sec:complexity}
We next assess the inference complexity of both components of the proposed approach: the \gls{nn}-based differential ambiguity estimator and the \gls{gd}-based solver. The computational cost is measured using the \gls{flop} count \cite{flop_motivation1_new, flop_motivation3_new, flop_motivation4_new, flop_motivation2_new, flop_count_reference, flop_count_reference_2}, where each elementary arithmetic operation (addition, subtraction, multiplication, division, square root) counts as one \gls{flop}, following the approach established in \cite{flop_count_reference, flop_count_reference_2, flop_motivation2_new}.

\subsection{Inference Complexity of the Proposed NN-based Differential Ambiguity Estimator}\label{subsec:nn_complexity}
The computational complexity of a fully connected layer with $n_i$ inputs and $n_o$ outputs is approximately $n_o(2n_i - 1) \approx 2n_o n_i$ \glspl{flop}, where bias additions and activation functions contribute negligibly to the overall cost for large layers. Since the \gls{nn}-based ambiguity estimation model described in Section~\ref{subsec:nn_model} consists exclusively of fully connected layers, the total inference complexity can be estimated as $ \mathcal{C}_{\text{NN}}\approx D^2(4\vert\mathcal{J}\vert+16)+D(2Q+4I)$ \glspl{flop}, where the computational overhead of the argmax operation is considered negligible and thus omitted from this analysis.

\subsection{Inference Complexity of the Proposed Gradient Descent-based Solver}\label{subsec:gd_complexity}
In each iteration, the distance calculations between the current \gls{ue} position and the selected \glspl{ap} involve $6 \vert \mathcal{J} \vert + 6$ \glspl{flop}, while the differential distance calculations consume $\vert \mathcal{J} \vert$ \glspl{flop} and the error calculations need $\vert \mathcal{J} \vert$ \glspl{flop}. The gradient calculation in \eqref{eq:gradient}, being the most computationally intensive step, demands $10\vert \mathcal{J} \vert$ \glspl{flop}, and updating the position estimation takes $4$ \glspl{flop}. In total, each iteration requires $18\vert \mathcal{J} \vert+10$ \glspl{flop}. For large $T$ values, where the final cost calculation and the threshold comparison are negligible, the computational complexity of the \gls{gd}-based solver is estimated as $\mathcal{C}_{\text{GD}}\approx T(18\vert \mathcal{J} \vert+10)$ \glspl{flop}.

\section{Numerical Results}
\subsection{Evaluation Assumptions and Parametrization}
We consider a $ 100 \, \text{m}^2 $ square evaluation region containing $I=9$ randomly distributed antenna points that emulate a \gls{cf} network topology. The \glspl{ap} are randomly deployed within the region while ensuring a minimum separation distance between them to avoid unrealistic clustering. Within this configuration, both the reference \gls{ap} and the indices of remaining \glspl{ap} are randomly designated. The uplink pilot design adheres to the 5G-NR \gls{srs} framework as specified in~\cite{3gpp_TS_38211}, operating at a carrier frequency of 2300\,MHz (5G NR band n40) with a subcarrier spacing of 15\,kHz. A total of 4 resource blocks are allocated for the pilot transmission, combined with a comb factor of 4, yielding an uplink reference signal bandwidth of 180\,kHz. A single reference symbol configuration~\cite{3gpp_TS_38211} is employed, and the modulation scheme used is BPSK. The channel is modeled using the free space path-loss model. Additional simulation parameters include a noise \gls{psd} of --174\,dBm/Hz, a receiver noise figure of 13\,dB, and uplink transmit power levels of $\{-20, -10, 0\}\,\text{dBm}$. 


The considered failure probabilities are $p_f\in \{0,10^{-3},10^{-2}\}$. We select the set of \gls{ap} indices for which differential ambiguities will be estimated as $\mathcal{J}=\{1,\dots,8\}$, such that the MLP-based model estimates all failure-independent differential ambiguities $\Delta\overline{\boldsymbol{z}} = \left[\Delta\overline{z}_{1},\dots,\Delta\overline{z}_{8}\right]^\top$. With the selected set of APs, the total number of possible output classes for the failure-independent differential ambiguities $\left[\Delta\overline{z}_{1},\dots,\Delta\overline{z}_{8}\right]^\top$ is calculated as $ Q=334 $. From a geometric perspective, estimating any two differential ambiguities yields two differential distance estimates which define two hyperbolas. The intersection of these hyperbolas provides already a unique positioning solution for the UE under ideal conditions. However, by estimating a larger set of differential ambiguities, the model benefits from additional geometric constraints, which enhance localization accuracy and improve robustness against noise and potential incorrect estimation of the differential ambiguities. The selection of the \gls{ap} indices whose ambiguities are estimated and the analysis of the performance-complexity trade-off as a function of $|\mathcal{J}|$ are left as important directions for future work. 


The layer parameter of the \gls{mlp}-based differential ambiguity estimator is set as $ D = 128 $. For evaluation purposes, separate \gls{nn} estimators are trained for each combination of transmission power level and failure probability $p_f$ using a supervised learning methodology. Each configuration utilizes $700 \times 10^3$ training samples and $150 \times 10^3$ validation samples, with randomly selected \gls{ue} positions for each sample instance. For each failure probability $p_f$, both the training and validation datasets are generated to reflect the corresponding failure conditions, ensuring that the model learns under the intended operating scenarios. The training process employs a batch size of 500 samples over 500 epochs. Network optimization is achieved through the Adam optimizer, configured with a learning rate of $10^{-3}$, while L2 regularization with a coefficient of $10^{-5}$, and a dropout rate of $0.1$ is applied during training to prevent overfitting. The \gls{gd}-based solver is configured to run for $T=500$ iterations, with a learning rate of $\alpha = 0.08$. 

\subsection{Differential Ambiguity Estimation Results}\label{subsec:ambiguity_results}
The performance of the \gls{mlp}-based differential ambiguity estimator is measured using the overall accuracy metric, defined as

\begin{equation}
    A_o = \frac{1}{S}\sum_{s=1}^{S} \mathbb{I}(\Delta\hat{\boldsymbol{z}}^s = \Delta\overline{\boldsymbol{z}}^s)\times 100\%,
\end{equation}   
where $\mathbb{I}(\cdot)$ is the indicator function, which equals 1 when its argument is true and 0 otherwise, and $\Delta\hat{\boldsymbol{z}}^s$ and $\Delta\overline{\boldsymbol{z}}^s$ are the estimated and true failure-independent differential ambiguity vectors for the test sample $s$, respectively. The obtained overall accuracy results for the trained models are summarized in Table~\ref{tab:ambiguity_estimation_performance}, based on a test set containing $S=100 \times 10^3$ samples. Evaluations are conducted across various uplink transmit power levels and \gls{ap} failure probabilities. For each value of the failure probability $p_f$, the test dataset is generated following the same procedure as for the training and validation datasets, ensuring that the failure patterns reflect the corresponding operating conditions.

As anticipated, $A_o$ improves consistently with increasing transmit power, reflecting the enhanced impact of signal quality on the estimator. With 9 distributed \glspl{ap} in the system, the probability of experiencing no failures is approximately $0.991$ for $p_f = 10^{-3}$. This indicates that the vast majority of test samples operate under failure-free conditions, yet a slight performance degradation is observed compared to the ideal case of $p_f=0$. This phenomenon occurs because the model encounters a small portion of samples with \gls{ap} failures at $p_f = 10^{-3}$, which may not provide sufficient training examples for the estimator to effectively learn robust prediction strategies under failure conditions. On the other hand, for $p_f = 10^{-2}$, approximately $8.65\%$ of the training samples experience at least one \gls{ap} failure. With this increased exposure to failure scenarios, the model learns to better handle degraded conditions, though a performance reduction compared to the failure-free case is still observed due to the inherent complexity of estimation under \gls{ap} failures. These results validate the robustness of the proposed ambiguity estimation model under varying \gls{ap} failure probabilities.

\begin{table}[t]
\centering
\caption{\textsc{Accuracy of differential ambiguity estimation}}
\label{tab:ambiguity_estimation_performance}
\begin{tabular}{l c c c c}
\toprule
& \multicolumn{3}{c}{\textbf{Transmit Power [dBm]}} \\  
\cmidrule(lr){2-4}
\textbf{Failure Probability} & --20 & --10 & 0 \\
\midrule
$p_f = 0$               & 98.69\% & 99.52\% & 99.80\% \\
$p_f = 10^{-3}$   & 98.15\% & 99.13\% & 99.38\% \\
$p_f = 10^{-2}$       & 96.93\% & 97.79\% & 98.11\% \\
\bottomrule
\end{tabular}
\end{table}

\subsection{Positioning Results}
To benchmark the positioning performance, we compare against the positioning approaches presented in \cite{ayten2025phase}, which assume ideal \gls{ap} operating conditions with $p_f=0$. Table~\ref{tab:positioning_errors} presents the 95th percentile positioning errors across different uplink transmit power levels and failure probabilities. The results demonstrate that the proposed approach achieves superior performance even under \gls{ap} failure scenarios. This performance advantage is attributed to the high accuracy of the \gls{mlp}-based ambiguity estimator, as evidenced in Table~\ref{tab:ambiguity_estimation_performance}.

\begin{figure}[t] 
    \centering
    \includegraphics[width=0.45\textwidth]{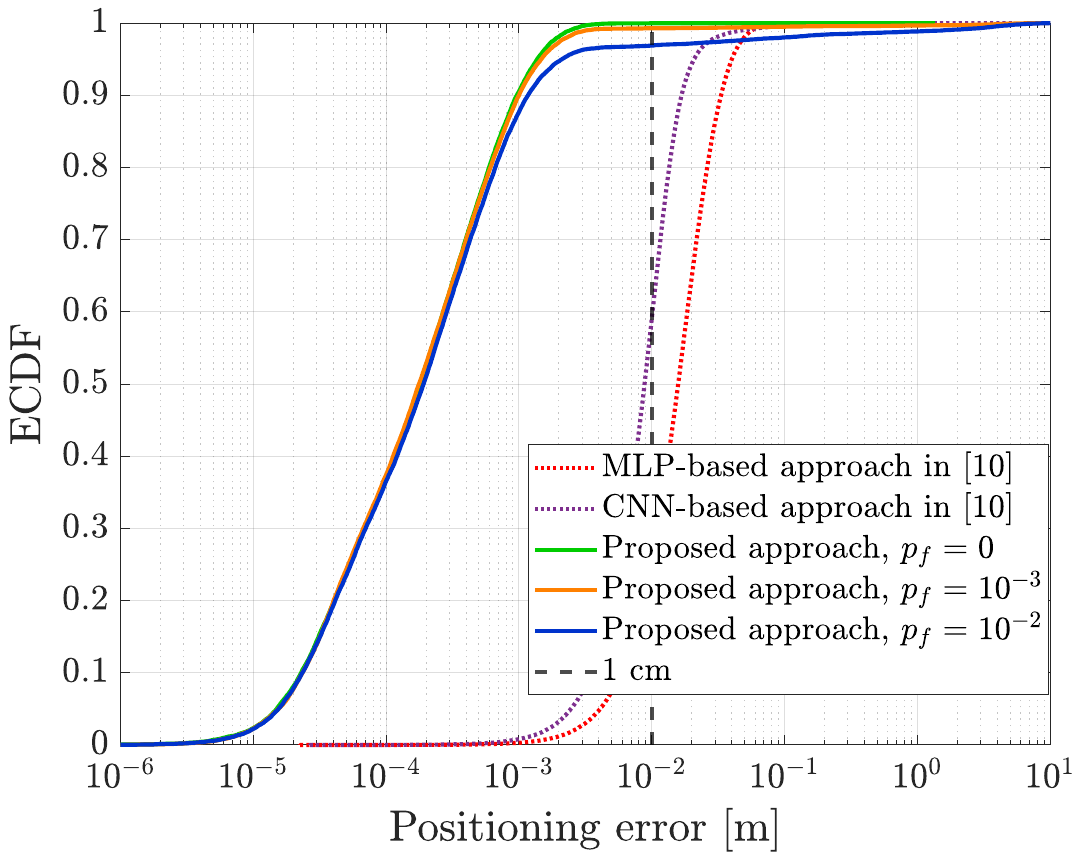}
    \caption{ECDF of the proposed approach and \gls{nn} approaches in previous work at a transmit power of 0\,dBm. The previous work benchmark performance assumes failure-free data ($p_f=0$).}\vspace{-0em}
    \label{fig:ecdf_figure}
\end{figure}

The \gls{ecdf} curves for a transmit power of 0~dBm are shown in Fig.~\ref{fig:ecdf_figure}. The proposed approach demonstrates superior accuracy in the sub-centimeter regime. The performance degradation under \gls{ap} failures can be explained through \eqref{eq:diff_dist_est}. Although the ambiguities are estimated with high accuracy as shown in Table~\ref{tab:ambiguity_estimation_performance}, \gls{ap} failures still impact \gls{ue} position estimation due to the addition of the estimated ambiguities with differential phase measurements, which are also affected by \gls{ap} failures \eqref{eq:phase_observations}. Nevertheless, these results confirm the reliability and precision of the proposed positioning approach under varying \gls{ap} failure conditions.

\begin{table}[t]
\centering
\caption{\textsc{95th percentile positioning errors of different models}}
\label{tab:positioning_errors}
\begin{tabular}{l c c c c}
\toprule
& \multicolumn{3}{c}{\textbf{Transmit Power [dBm]}} \\  
\cmidrule(lr){2-4}
\textbf{Approach}  & --20 & --10 & 0 \\
\midrule
Proposed approach, $p_f=0$          & 1.35cm & 0.42cm & 0.14cm \\
Proposed approach, $p_f=10^{-3}$    & 1.43cm & 0.45cm & 0.15cm \\
Proposed approach, $p_f=10^{-2}$    & 1.88cm & 0.65cm & 0.21cm\\
MLP-based approach in \cite{ayten2025phase}, $p_f=0$       & 4.54cm & 4.05cm & 4.01cm \\
CNN-based approach in \cite{ayten2025phase}, $p_f=0$        & 3.16cm & 2.83cm & 2.08cm \\
\bottomrule
\end{tabular}
\end{table}

\subsection{Failure Detection Results}\label{subsec:failure_detection_results}
In order to evaluate the failure detection capabilities of the proposed models, four distinct test sets were constructed, each consisting of $S=100 \times 10^3$ samples. The first test set, referred to as the No Failure dataset, contains samples in which no APs are malfunctioning. In contrast, the remaining three datasets—1 Failure, 2 Failures, and 3 Failures—simulate the scenarios in which one, two, or three APs, respectively, are always failing. The indices of the failing APs in these test sets are selected randomly for each test sample to ensure generality and prevent bias toward any specific APs. The failure detection performance of the models under these varying failure conditions is assessed by calculating the ratio of test samples that successfully pass the threshold test, i.e., $E_T\leq \tau $, with the detection threshold set here to $\tau = 10^{-4}$. 

The failure detection results are presented in Table~\ref{tab:threshold_test_1e_5} for models trained with varying failure probabilities and transmit power levels. The model trained with the failure probability of $p = 0$, which is trained exclusively on No Failure samples, demonstrates the highest pass rate on the No Failure test set. It consistently outperforms models trained with $p = 10^{-3}$ and $p = 10^{-2}$. On the other hand, in the presence of failures in the test data, a different trend emerges. As the value of $p$ increases, indicating more exposure to failure cases during training, the models demonstrate worse detection capability by allowing a larger portion of the samples to pass the threshold test. Notably, for all three model configurations, the percentage of samples that pass the threshold in the No Failure test set exceeds $99\%$, indicating strong agreement with non-faulty data. In contrast, for the 3 Failures test set, the pass ratio drops below $1\%$ across all models. This stark contrast in detection rates between non-failure and multiple-failure scenarios highlights the models’ ability to distinguish between nominal and anomalous conditions. In general, the high failure detection reliability contributes to increased positioning integrity, particularly when it comes to the ECDF tails visible in Fig.~\ref{fig:ecdf_figure}.

\begin{table}[t]
\centering
\caption{\textsc{Ratio of The Test Samples that Passes the Threshold Test, $\tau=10^{-4}$}}
\label{tab:threshold_test_1e_5}
\begin{tabular}{@{}l c c c c c@{}}
\toprule
& & \multicolumn{3}{c}{\textbf{Transmit Power [dBm]}} \\  
\cmidrule(lr){3-5}
\textbf{Testset} & \textbf{$p_f$ in the training set}  & --20 & --10 & 0 \\
\midrule
\multirow{4}{*}{No failure}
& $0$       & 99.70\% & 99.71\% & 99.83\% &  \\
& $10^{-3}$ & 99.49\% & 99.62\% & 99.75\% &  \\
& $10^{-2}$ & 99.46\% & 99.54\% & 99.68\% &  \\
\midrule
\multirow{4}{*}{1 failure}
& $0$       & 16.05\% & 17.01\% & 17.19\% &  \\
& $10^{-3}$ & 16.97\% & 17.29\% & 17.90\% &  \\
& $10^{-2}$ & 17.08\% & 18.53\% & 18.97\% &  \\
\midrule
\multirow{4}{*}{2 failures}
& $0$       & 2.11\% & 2.48\% & 2.69\% &  \\
& $10^{-3}$ & 2.14\% & 2.60\% & 2.88\% &  \\
& $10^{-2}$ & 2.26\% & 2.82\% & 2.93\% &  \\
\midrule
\multirow{4}{*}{3 failures}
& $0$       & 0.11\% & 0.15\% & 0.47\% &  \\
& $10^{-3}$ & 0.14\% & 0.36\% & 0.53\% &  \\
& $10^{-2}$ & 0.42\% & 0.50\% & 0.65\% &  \\
\bottomrule
\end{tabular}
\end{table}

\subsection{Inference Complexity Comparison}
Based on the inference complexity derivations in Section~\ref{sec:complexity}, the \gls{nn}-based differential ambiguity estimator has a complexity of $\mathcal{C}_{\text{NN}}\approx 0.877 \times 10^6$ \glspl{flop}, while the \gls{gd} solver requires $\mathcal{C}_{\text{GD}}\approx 0.077 \times 10^6$ \glspl{flop}. The combined approach yields a total computational complexity of $\mathcal{C}_{\text{T}}\approx 0.954 \times 10^6$ \glspl{flop}, achieving complexity reductions of approximately $30.7\%$ and $16.9\%$ relative to \gls{mlp}-based positioning ($1.376 \times 10^6$ \glspl{flop}) and \gls{cnn}-based positioning ($1.148 \times 10^6$ \glspl{flop}) methods reported in \cite{ayten2025phase}, respectively.

\section{Conclusion}
This work proposed methods to improve the accuracy of carrier phase-only UE positioning in distributed MIMO or cell-free systems beyond the existing state-of-the-art, while also considering and involving practical antenna element failures by leveraging a learning-based \gls{nn} approach. The proposed differential ambiguity estimator achieves high accuracy across varying uplink transmit powers and failure probabilities, demonstrating robustness under practical system impairments. Positioning performance remains highly reliable, with 95th percentile errors consistently below $2$ cm—even under moderate antenna failure rates—validating the method's suitability for stringent centimeter-level positioning requirements. Furthermore, the failure detection capability of the models shows strong generalization: training with failure scenarios improves robustness, enabling effective identification of faulty antenna conditions without compromising performance in fault-free cases. Overall, the results confirm that deep learning-driven phase-only positioning offers a promising solution that is resilient to antenna failures, ensuring precise and dependable localization in next-generation mobile networks. Future work will focus on investigating the impact of \gls{nlos} and multipath effects on the proposed positioning framework. Additionally, we aim to explore the influence of residual phase-calibration errors at the network side.

\section{ACKNOWLEDGEMENT}
This work was supported by the MiFuture project under the HORIZON-MSCA-2022-DN-01 call (Grant number: 101119643), and by the SNS JU project 6G-DISAC under the EU's Horizon Europe research and innovation Program under Grant Agreement No 101139130. The work was also  supported in part by Business Finland under the 6G-ISAC project, and in part by the Research Council of Finland under the grant 359095.

\balance
\bibliographystyle{IEEEtran}
\bibliography{References}

\end{document}